\documentclass[useAMS,usenatbib]{mn2e}
\usepackage{epsfig,amsmath,natbib}
\usepackage{color}

\newcommand{\be}{\begin{equation}} 
\newcommand{\e}{\end{equation}} 
\newcommand{\bear}{\begin{eqnarray}} 
\newcommand{\ear}{\end{eqnarray}}

\def\be{\begin{equation}} 
\def\ee{\end{equation}}

\def\gsim{\lower.5ex\hbox{\gtsima}} 
\def\lsim{\lower.5ex\hbox{\ltsima}} 
\def\gtsima{$\; \buildrel > \over \sim \;$} 
\def\ltsima{$\; \buildrel < \over \sim \;$} 
\def\prosima{$\; \buildrel \propto \over \sim \;$} 
\def\gsim{\lower.5ex\hbox{\gtsima}} 
\def\lsim{\lower.5ex\hbox{\ltsima}} 
\def\simgt{\lower.5ex\hbox{\gtsima}} 
\def\simlt{\lower.5ex\hbox{\ltsima}} 
\def\simpr{\lower.5ex\hbox{\prosima}}

\title[CO(17--16) line in a quasar at $z>6$] 
{First CO(17--16) emission line detected in a $z>6$ quasar}

\author[S. Gallerani, A. Ferrara, R. Neri, R. Maiolino]{ 
S. Gallerani$^{1}$, A. Ferrara$^{1}$, R. Neri$^{2}$, R. Maiolino$^{3,4}$ \\ 
$^1$ Scuola Normale Superiore, Piazza dei Cavalieri 7, 56126, Pisa, Italy \\ 
$^2$ IRAM, 300 Rue de la Piscine, 38406 St--Martin--d' H$\rm \grave{e}$res, France\\
$^3$ Cavendish Laboratory, University of Cambridge, 19 J. J. Thomson Ave., Cambridge CB3 0HE, United Kingdom\\
$^4$ Kavli Institute for Cosmology, University of Cambridge, Madingley Road, Cambridge CB3 0HA, United Kingdom 
} 
\date{\today} 
\pagerange{\pageref{firstpage}--\pageref{lastpage}} \pubyear{2014} 
\begin{document} 
\maketitle 
\label{firstpage} 
\begin{abstract} 
We report the serendipitous detection of the CO(17--16) emission line toward the quasar SDSS\,J114816.64+525150.3 (J1148) at redshift $z\simeq 6.4$ obtained with the Plateau de Bure Interferometer. The CO(17--16) line is possibly contaminated by OH$^+$ emission, that may account for $\sim 35-60$\% of the total flux observed. Photo-Dissociation and X-ray Dominated Regions (PDRs and XDRs) models show that PDRs alone cannot reproduce the high luminosity of the CO(17-16) line relative to low-J CO transitions and that XDRs are required. By adopting a composite PDR+XDR model we derive molecular cloud and radiation field properties in the nuclear region of J1148. Our results show that highly excited CO lines represent a sensitive and possibly unique tool to infer the presence of X-ray faint or obscured supermassive black hole progenitors in high-$z$ galaxies.
\end{abstract} 
\section{Introduction} 
One of the major observational breakthroughs of the last decade has been the discovery of a large sample of quasars at $z>6$ (Fan 2006), with the current record holder being quasar ULAS J1120+0641 at $z = 7.085$ (Mortlock et al. 2011). These observations have unquestionably shown that the most distant and luminous quasars are powered by gas accretion onto $\gsim 10^9 M_{\odot}$ black holes (Jiang et al. 2007; Kurk et al. 2007, Kurk et al. 2009).\\
Such evidence poses thorny questions on the formation of these supermassive black holes (SMBHs) and on the detection and characterization of their lower mass ($10^{6-8} ~{\rm M_{\odot}}$) ancestors (e.g. Tanaka \& Haiman 2009; Volonteri 2010, Petri, Ferrara \& Salvaterra 2012). In fact, since at $z>6$ the Universe is less than 1 Gyr old, it is difficult to understand how SMBHs have formed over such a short time scale. In hierarchical galaxy evolution models, BH growth can occur through mergers of nuclear BHs and by gas accretion. Under the assumption that SMBH progenitors are accreting no faster than the Eddington rate on small seeds ($< 10^4 M_{\odot}$), the latter must appear at very high redshift ($z>10$). SMBH ancestors have yet to be detected, despite being predicted and extensively searched for (e.g. Cowie et al. 2013). One interesting possibility is that ancestors are enshrouded by a thick gas and dust cocoon heavily absorbing their high energy radiation, therefore eluding optical/X-ray observations.\\
The availability of sensitive millimeter interferometers allows the detection of early quasars and galaxies through far-infrared emission lines (e.g. Maiolino et al. 2005; Riechers et al. 2009; Walter et al. 2009; Maiolino et al. 2012; Nagao et al. 2012; Neri et al. 2014). In this context, the advantage of FIR lines is that they are unaffected by dust extinction, therefore allowing the detection even of high-$z$ dust obscured sources (e.g. Gallerani et al. 2012). \\
Here we present Plateau de Bure Interferometer (PdBI) observations of the quasar SDSS\,J114816.64 +525150.3 (hereafter J1148) at redshift $z\sim 6.4$. J1148 is one of the most distant and luminous ($M_{\rm 1450}\sim -27.82$) known quasars (Fan et al. 2003; White et al. 2003), therefore representing an ideal laboratory to study early galaxy/star formation. Near infrared observations have shown that this luminous quasar is powered by a SMBH of mass $M_{\bullet}=3.0\pm 2.5\times10^9$ M$_{\odot}$ (Willott et al. 2003). The interstellar medium (ISM) of J1148 has been studied through several CO rotational transitions: CO(1--O); CO(3--2); CO(6--5); CO(7--6) (Bertoldi et al. 2003a; Walter et al. 2003; Riechers et al. 2009). In this paper, we present the discovery of the CO(17--16) line.\\
CO lines have been detected so far in local ($z<0.1$) luminous infrared and starburst galaxies up to $J_{\rm max}=13$ (Greve et al. 2014; Meijerink et al. 2013; Panuzzo et al. 2010), and in a Seyfert 2 galaxy up to $J_{\rm max}=30$ (Hailey-Dunsheath et al. 2012). At higher redshifts, CO detections are limited to $J_{\rm max}\leq 11$ transitions, both in $2\leq z\leq 5$  (e.g. Omont et al. 1996, Barvainis et al. 1997; Weiss et al., 2007; Ao et al. 2008) and in $z\sim 6$ (e.g. Wang et al. 2010; Riechers et al. 2013) quasars and galaxies (see also the reviews by Carilli \& Walter 2013 and Casey et al. 2014). 
Given the above mentioned results, the CO(17--16) line detected in J1148 at $z=6.4$ represents the most excited CO rotational transition in the millimeter range ever detected not only in $z\sim 6$ quasars, but also in $z>0.1$ galaxies. 
\begin{figure*}
   \begin{center} 
   \psfig{figure=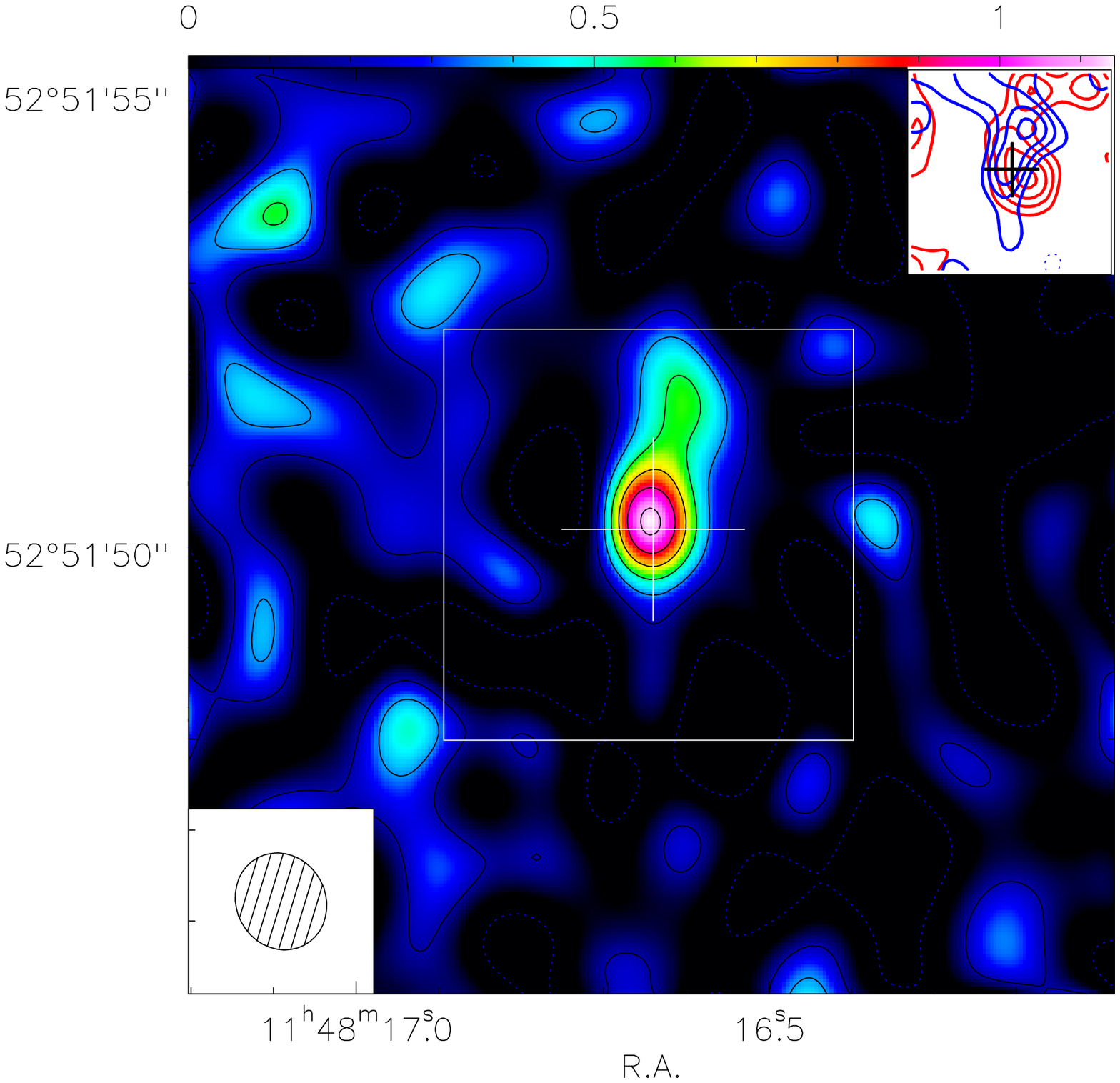,width=83mm,clip=t,angle=0}
\psfig{figure=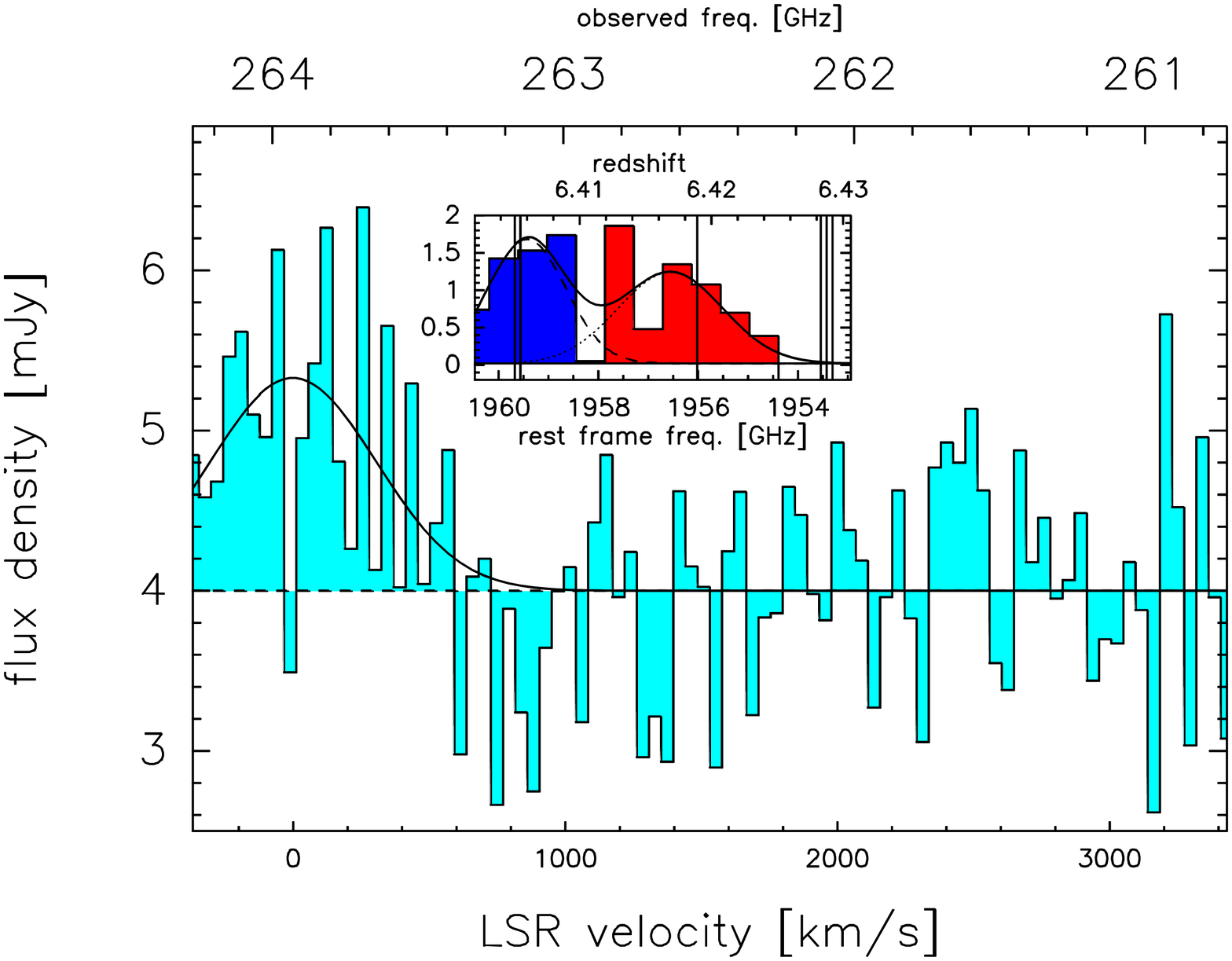,width=93mm,clip=t,angle=0}
   \end{center}
\caption{
{\em Left panel: } PdBI CO(17--16) emission detected in J1148. The 1$\sigma$ noise in the images is 0.183\,mJy\,beam$^{-1}$ and contours are plotted from -1$\sigma$ to 6$\sigma$. The cross indicates the optical position of J1148. The observations were taken with the PdBI in the C+D configuration, yielding a resolution of $1.1''\!\times\!0.98''$; the beam is plotted in the lower left of the panel).
 {\em Right: } Spectrum of the CO(17--16) line of J1148 as observed with the PdBI (channel width: 39\,MHz\,=\,44\,km\,s$^{-1}$, noise per channel: 0.8\,mJy), shown on the top of a 4.0$\pm$0.1~mJy continuum emission at 262 GHz. A Gaussian fit to the PdBI data gives a velocity full width at half maximum of FWHM=720$\pm$217\,km\,s$^{-1}$, a peak flux of $F_{\rm peak}$=1.33$\pm$0.43~mJy, and a redshift of $z=6.410\pm 0.005$, assuming the rest frame frequency of the CO(17--16) emission line ($\nu_{\rm RF}=1956.018137$~GHz). See the text for the description of the insets in both panels.
}
\label{fig1}
\end{figure*}
\section{Observations}\label{Obs} 
Continuum observations at 262 GHz of J1148 were performed using the IRAM Plateau de Bure interferometer (PdBI) with 6 antennas in May 2010 (D configuration) and 5-6 antennas in Nov-Dec 2012 (C configuration). The beam size resulting from the C+D configuration using natural weighting is $1.1''\times 0.98''$, with a position angle of 27$^{\rm o}$. The on-source observing time is 19.1 hr (6-antenna equivalent). We use the quasar 1150+497 for phase and amplitude calibration. The absolute flux calibration was derived by observing several sources, namely: MWC349, 3C84, 0923+392, 3C279, 0851+202, 3C273, 1055+018, 1144+542, 3C345. The setting of the PdBI receivers covered a frequency range $260.758\leq \nu_{\rm obs}/\rm GHz \leq 264.274$.\\
We serendipitously detected strong line emission with high significance (6.2$\sigma$) in our data, perfectly centered at the SDSS location of this source, at an observed frequency $263\lsim \nu_{\rm obs}/\rm GHz\lsim 264$ (Fig. 1). We estimate the continuum at 262 GHz through the line-free channels (i.e. $v>800~\rm km~s^{-1}$, $\Delta v= 2800~\rm km~s^{-1}$) of the total 3.5~GHz wide spectrum. We obtain a $1\sigma$ sensitivity of $0.08$~mJy~beam$^{-1}$ and a continuum flux of $F_{cont}=4.0\pm 0.1$~mJy.\\
Given the most accurate redshift determination ($z=6.4189\pm 0.0006$) for J1148 from the [CII] line (Maiolino et al. 2005), the frequency range in which the emission line is detected corresponds to $1951\lsim \nu_{\rm RF}/\rm GHz \lsim 1959$, in the rest frame of the source. The CO(17--16) emission line ($\nu_{\rm RF}=1956.018137$~GHz), as well as five $\rm OH^+$ rotational transitions  ($1953.426326\leq \nu_{\rm RF}/\rm GHz\leq 1959.675435$), fall in this frequency interval. This line complex has been recently observed in NGC1068, a nearby Seyfert 2 galaxy (Hailey-Dunsheath et al. 2012). According to that study, the combined line profile is largely ($\sim 65\%$) dominated by the CO(17--16) line. Therefore, we first assume the rest frame frequency of the CO(17--16) emission line and we fit the PdBI data with a Gaussian function. We get a velocity full width at half maximum of FWHM=720$\pm$217\,km\,s$^{-1}$, a peak flux of $F_{\rm peak}$=1.33$\pm$0.43~mJy, and a redshift of $z=6.410\pm 0.005$. This converts into a total velocity--integrated flux S$\Delta v=$1.01$\pm 0.16$ Jy\,km\,s$^{-1}$.\\ 
Compared to lower $J$ CO transitions and the [CII] line in J1148+52, the CO(17-16) line is both blue-shifted ($v=-416\pm 233$~km~s$^{-1}$) and broader ($\rm 720 \pm 270$ km s$^{-1}$ vs. $\rm 297 \pm 35$ km s$^{-1}$ of CO(7-6)). Although these differences are marginal (significance 1.8 and 1.6$\sigma$, respectively), they can be explained by the possible contamination from $\rm OH^+$ emission. In fact, our data can be very well described through a double Gaussian. In the inset in the right panel of Fig. 1, we show the spectrum rebinned to 88 km~s$^{-1}$. The dotted line (shaded red region) represents the CO(17-16) line ($z_{\rm CO(17-16)}\sim 6.418$; FWHM$_{\rm CO(17-16)}\sim 297$\,km\,s$^{-1}$), while the dashed line (blue shaded region) denotes the OH$^+$ line ($z_{\rm OH^+}\sim 6.420$; FWHM$_{\rm OH^+}\sim 373$\,km\,s$^{-1}$). The quality of our data is not high enough to determine the errors on the 6 free parameters of the fit, above all because the emission is not centered in our observational set-up. However, it is reassuring to find that through the double Gaussian model the properties of the CO(17--16) and OH+ lines (inferred redshift and FWHM) are much closer to the expected ones. Moreover, in this way we can quantify the relative contribution of the CO(17--16) line to the total emission, that results to be $\sim 40$\%.\\
We further compute the map of the CO(17--16) and OH$^+$ lines by integrating over the channels denoted by the red and blue shaded regions, respectively. In the inset in the left panel of Fig. 1 (on the top right), we show the 1$\sigma$ levels of the CO(17--16) and OH$^+$ emission through the red and blue lines, respectively (1$\sigma_{\rm CO(17-16)}=0.09$~Jy~km~s$^{-1}$beam$^{-1}$; 1$\sigma_{\rm OH^+}=0.07$~Jy~km~s$^{-1}$beam$^{-1}$, dotted blue lines represent -3$\sigma$ levels from the OH$^+$ map). We detect both emissions at 4.3$\sigma$. By integrating over the 3$\sigma$ contours of both maps, we find that the total emission we detect is $\sim 50$\% due to the CO(17-16) transitions.\\
To summarize, the CO(17--16) line is contributing $\sim 40-65$\% to the emission we detect. By conservatively considering 40\% as our fiducial value, the total integrated flux S$\Delta v$ we observe corresponds\footnote{We use $H_0\!=\!67.3$\,km\,s$^{-1}\,$Mpc$^{-1}$, $\Omega_{\Lambda}=0.685$ and $\Omega_{m}=0.315$ throughout (Planck Collaboration 2013).} to a CO(17--16) luminosity $L_{\rm CO(17-16)}=(4.9\pm 1.1)\times 10^8~L_{\odot}$ (Solomon et al. 1992).
\section{CO SLED}
\begin{figure}
   \begin{center} 
   \psfig{figure=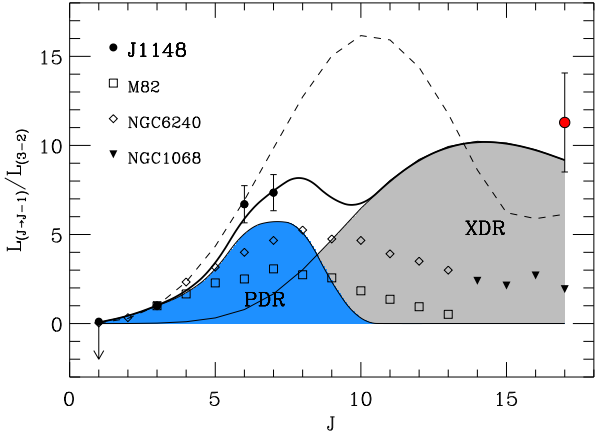,width=85mm,clip=t,angle=0}
   \end{center}
\caption{CO Spectral Line Energy Distribution (COSLED) of J1148. The circles represent data for the following CO transitions: CO(1--0) (upper limit from Bertoldi et al. 2003a); CO(3--2) (Walter 2003); CO(6--5) (Bertoldi 2003a); CO(7--6) (Reichers 2009); CO(17--16) (This work). The dashed line denotes the best-fitting PDR model, while the thick solid line shows the best-fitting composite model to our data. 
In the latter case, the relative contributions from PDR and XDR models alone are also shown by the blue and gray shaded regions, respectively. We also plot the COSLEDs observed in several local galaxies:  M82 (empty squares; Panuzzo et al. 2010); NGC6240 (empty diamonds; Meijerink et al. 2013); NGC1068 (filled triangle; Hailey-Dunsheath et al. 2012).}
\label{fig2}
\end{figure}
The CO Spectral Line Energy Distribution (COSLED) is defined as the ratio of the ($J \rightarrow J-1$) CO rotational transition luminosity to a fixed CO transition\footnote{We normalize the COSLED to the CO(3--2) transition.}, as a function of $J$. The COSLED is considered a sensitive probe of the molecular gas kinetic temperature, $T_K$, in turn controlled by radiative heating from stars and AGN (e.g. Obreschkow et al. 2009). More specifically, a higher $T_K$ boosts higher $J$ transitions, thus pushing the COSLED maximum towards larger $J$  values. The presence of high energy ($\geq 1~{\rm keV}$) photons emitted by an AGN causes the CO line intensities to rise well beyond $J=10$, making these highly excited lines a tracer of quasar activity and a powerful diagnostics of X-ray Dominated Regions (XDR) (e.g. Schleicher et al. 2010). 

In Fig.~2, we show the COSLED observed in J1148 through filled circles. We adopt available radiative transfer results (Meijerink et al. 2005; 2007) to compute the COSLED predicted by Photo-Dissociation Regions (PDR, i.e. excited by UV radiation field) and XDR models. We explore different gas densities ($n_{\rm PDR}$, $n_{\rm XDR}$ in ${\rm cm^{-3}}$), UV fluxes ($G_{\rm 0}$ in Habing units of ${\rm 1.6 \times 10^{-3} erg~s^{-1}cm^{-2} }$), and X-ray fluxes ($F_{\rm X}$ in ${\rm erg~s^{-1}cm^{-2}}$), at the cloud surface. 

First, we try to reproduce the observed COSLED with pure PDR models. In this case, the best fitting model ($n_{\rm PDR}=10^{6}~\rm cm^{-3}$, $G_{\rm 0}=10^{4}$, dashed line in Fig.~2). This result clearly shows that PDR models fail to reproduce our high-J (J$\geq$7) CO line observations. In fact, in this case, we obtain $\chi^2_{\rm BF}=9.5$, namely P$(\chi^2>\chi^2_{\rm BF})=0.02$. 
Therefore, we consider a slightly more complex but physically plausible composite model, in which the molecular cloud (MC) emission is the sum of the contribution from a higher-density XDR region embedded in a more rarefied and extended PDR envelope. The solid line in Fig. 2 represents the best-fitting composite model to available data. In this case, we obtain $\chi^2_{\rm BF}=0.9$, i.e. P$(\chi^2>\chi^2_{\rm BF})=0.6$. According to this model, individual MCs have a typical mass $M_{\rm c}\sim 2\times 10^5~{\rm M_{\odot}}$ and a radius $r_{\rm c} \sim 10$ pc; the XDR core ($n_{\rm XDR}=10^{4.25\pm 0.25}$) is irradiated by an X-ray flux $F_{\rm X}=160\pm 70$, while the FUV flux at the PDR surface ($n_{\rm PDR}=10^{3.25\pm 0.25}$) is $G_{\rm 0}=10^{4.0\pm 0.25}$. The quoted errors must be considered as upper limits on the actual 1$\sigma$ values, since the grids of PDR and XDR models are not sufficiently fine-grained to correctly determine the errors on the best fit model parameters. Moreover, the limited number of CO lines detected so far prevents us to fully characterize the uncertainties/degeneracies of the model parameters. Additional observations of high-J CO lines ($7<J<17$), more accurate XDR and PDR models, as well as improvements of XDR diagnostics (e.g. OH$^+$) will be crucial to further test our predictions.\\
\section{Observed X-ray flux}
\begin{figure}
   \begin{center} 
   \psfig{figure=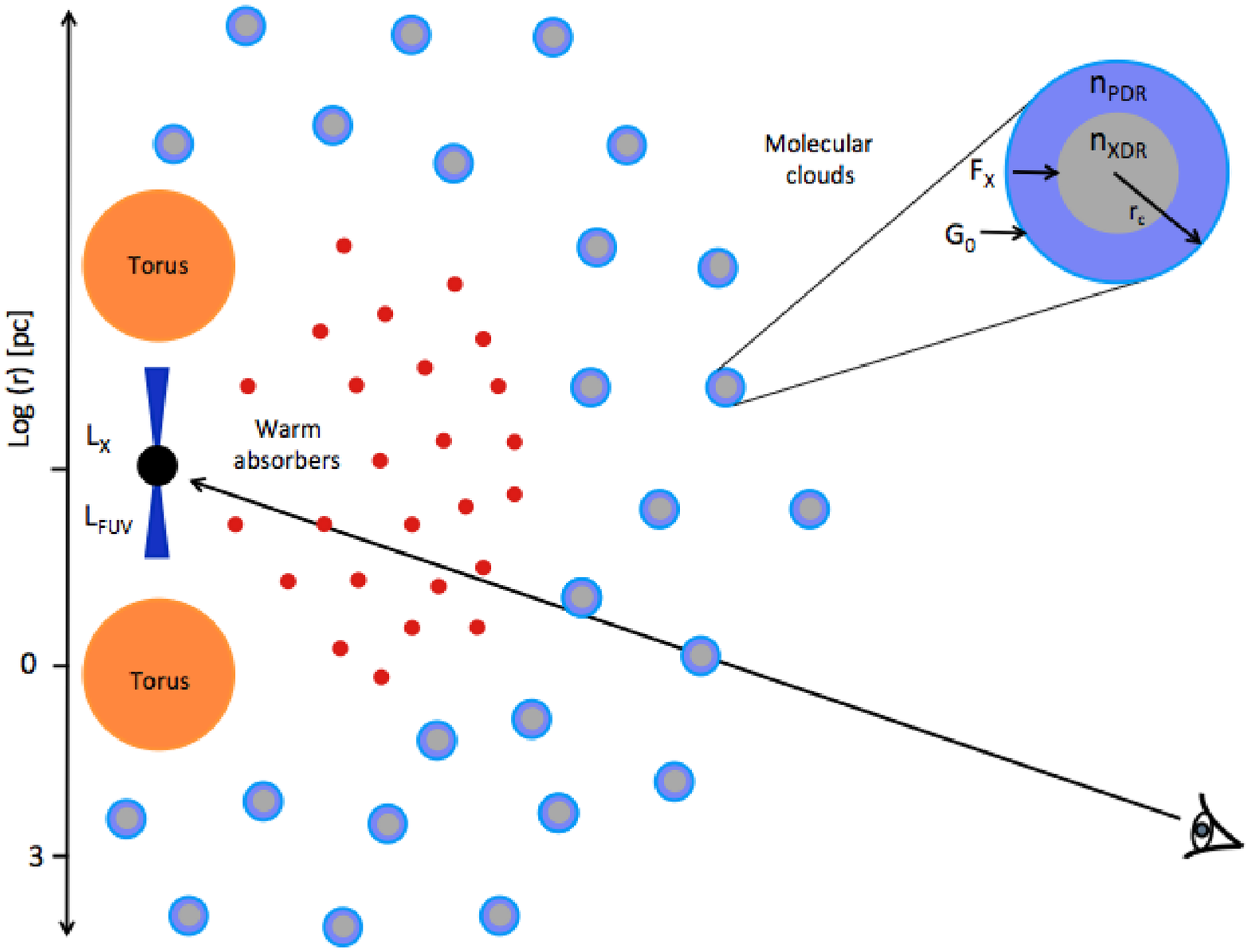,width=70mm,clip=t,angle=90}
   \end{center}
\caption{Schematic representation of the gas distribution in J1148.}
\label{toymodel}
\end{figure}
High-resolution, spatially resolved, VLA observations of the CO(3--2) line have discovered a molecular component of total mass $M_{\rm H_2}$=2.2$\times10^{9}$\,M$_\odot$, extending up to a radius $r_{\rm H_2}= 2.5$~kpc from the center (Walter et al. 2004). By combining the results of our analysis with these observations, we can build up a simple model of the MCs in J1148 (see Fig. \ref{toymodel}). According to this model, the total numer of clouds is $N_{c}=M_{\rm H_2}/M_{\rm c}$, their covering filling factor $f_{\rm c}=N_{\rm c}(r_c/2r_{\rm H_2})^2$, and the number of clouds intercepted by the line of sight towards J1148 is $N_{c}^{||}=3f_{\rm c}$.

As X-ray observations of J1148 are not yet available, we can use the properties of J1148 MCs resulting from our modelling to predict its flux in the Chandra soft X-ray band ($0.5\leq E/{\rm keV}\leq 2~$): 
\begin{equation}
F_{\rm X}^{\rm soft}=~F_{\rm X}^{\rm 0}~\exp(-\tau_{\rm tot})=~F_{\rm X}^{\rm 0}~\exp[-(\tau_{\rm abs}+\tau_{\rm MC})], 
\end{equation}
where $F_{\rm X}^{\rm 0}$ is the intrinsic (unabsorbed) X-ray flux integrated between $3.7\leq E/{\rm keV}\leq 14.8$, $\tau_{\rm MC}$ is the total optical depth of MCs along the los  to X-ray photons, and $\tau_{\rm abs}$ is the optical depth of absorbers eventually intervening along the line of sight between the black hole and the clouds. We do not take into account the torus contribution to the total optical depth since J1148 is a Type I quasar (Juarez et al. 2009).
 
We compute $F_{\rm X}^{\rm 0}$ through the following equation:
\begin{equation}
F_{\rm X}^{\rm 0}=\frac{(1+z)~L_{\rm X}}{4\pi d_{\rm L}(z)^2}, 
\end{equation}
where $L_{\rm X}=f_{\rm X}L_{\rm E}$, $f_{\rm X}=0.007$ is the X-ray bolometric correction (Hopkins et al. 2007), $L_{\rm E}=1.26\times 10^{\rm 38}~(M_{\bullet}/M_{\odot})$ erg~s$^{-1}$ is the Eddington luminosity, $d_{\rm L}(z)$ is the luminosity distance at the J1148 redshift.

To compute $\tau_{\rm abs}$, we need to solve the following set of equations:
\begin{eqnarray}
G_0&=&\frac{L_{\rm FUV}}{4\pi d_c^2}\exp[-\sigma_{FUV}N_H^{abs}](1+\frac{G_0^*}{G_0^{\rm QSO}})\\
F_X&=&\frac{L_X}{4\pi d_c^2}\exp[-\tau_{\rm X,RF}(N_{\rm H}^{\rm abs})+\tau_{\rm X,RF}(N_{\rm H}^{\rm PDR})]\\
\tau_{\rm X}&=&\tau_{\rm X}(N_{\rm H})
\end{eqnarray}
where $L_{\rm FUV}$ is the (FUV) luminosity in the Habing band (6--13.6 eV), obtained by assuming a quasar spectrum template (Telfer et al. 2002) and bolometric corrections, $d_c$ is the (unknown) cloud typical distance from the black hole, $\sigma_{\rm FUV}$ is the PDR optical depth to FUV photons, $N_{\rm H}^{\rm abs}$ is the (unknown) absorbers column density, $\tau_{\rm X,RF}$ is the optical depth to X-ray photons with energies ($1\leq {\rm E/keV}\leq 100$), and $G_{\rm 0}^*$~($G_{\rm 0}^{\rm QSO}$), is the stellar (quasar) contribution to the FUV radiation field\footnote{We assume $G_{\rm 0}^*=0.5 G_{\rm 0}^{\rm QSO}$ (Walter et al. 2009), but our results are completely insensitive to this choice}.
For $\tau_{\rm c}$, we use the following equation: $\tau_{\rm c}=\tau_{\rm X,obs}(N_{\rm H}^{\rm c})$, where $\tau_{\rm X,obs}$ is the optical depth\footnote{The computations of the photoelectric optical depth are based on the calculations presented by Morrison \& McCammon (1983).} to the rest frame photons ($3.7\leq E/{\rm keV}\leq 14.8$) which at the J1148 redshift are shifted in the Chandra soft band ($0.5\leq E/{\rm keV}\leq 2$), and $N_{\rm H}^{\rm c}=2~[n_{\rm XDR}r_{\rm XDR}+n_{\rm PDR}(r_{\rm c}-r_{\rm XDR})]$ is the column density of each cloud.  Finally, the total MC optical depth is given by $\tau_{\rm MC}=N_{c}^{||}~\tau_{\rm c}$. 
We find that $\tau_{\rm abs}=0.01$ is negligible with respect to $\tau_{\rm MC}\sim 1.1$. In fact,
our model predicts that the intervening material between the black hole and the MCs has a relatively low hydrogen column density ($N_{\rm H, abs} = 3\times 10^{20}$ cm$^{-2}$) with respect to MCs intercepted along the line of sight  ($N_{\rm H, MC} = 4\times 10^{23}$ cm$^{-2}$).  
In conclusion, for J1148, we expect a flux $F_{\rm X}^{\rm soft}=1.2\times 10^{-14}~{\rm erg~s^{-1}~cm^{-2}}$ in the soft Chandra band. By performing similar calculations for the hard ($2\leq E/{\rm keV}\leq 8$) band, we obtain $F_{\rm X}^{\rm hard}=3.5\times 10^{-14}~{\rm erg~s^{-1}~cm^{-2}}$. 
\section{SMBH progenitors detectability}
The next step is to understand whether we can use the CO(17--16) line to identify high-$z$ quasars that cannot be detected in X-rays. 
We consider a scaled-down version of J1148, namely a $z=7$ quasar powered by a black hole of mass $10^3$ times lower than J1148 ($M_{\bullet}=3\times 10^{6}~{\rm M_{\odot}}$). We assume that the surrounding molecular gas mass is decreased by the same factor ($M_{\rm H_2}=2.2\times 10^7~M_{\odot}$) and that the quasar follows the $M_{\bullet}-\sigma_*$ relation observed in the local Universe (Ferrarese \& Merritt 2000; Gebhardt et al. 2000). This implies a velocity dispersion $v_c=80~\rm km~s^{-1}$, which corresponds at $z=7$ to a dark matter halo mass $M_h=1.5\times 10^{10}~M_{\odot}$ and to a virial radius $r_{vir}=10~\rm kpc$. Assuming angular momentum conservation, we then distribute the molecular gas in a radius $r_{\rm H_2}= (\lambda/\sqrt 2) r_{\rm vir}=0.28~\rm kpc$, where $\lambda=0.04$ is a typical halo spin parameter.\\
Albeit predicted and extensively searched, objects of this type have not been detected so far, most likely because they are enshrouded by a thick (column density $N_{\rm H}^*>10^{24}~\rm cm^{-2}$) gas and dust cocoon heavily absorbing their optical/X-ray radiation (Comastri 2004). If so, undetected SMBH progenitors must have $N_{\rm H}>N_{\rm H}^*$. We consider the subset of the above PDR+XDR models satisfying this condition and compute their expected CO(17--16) luminosity through the following equation: 
\begin{equation}
L_{\rm CO(17-16)}=\frac{1}{2} N_{\rm c} \Omega_{OA} r_{\rm c}^2 (F_{\rm CO(17-16)}^{PDR}+F_{\rm CO(17-16)}^{XDR}), 
\end{equation}
where $N_{\rm c}$ is the number of clouds intercepted along the line of sight, $F_{\rm CO(17-16)}^{PDR}$ ($F_{\rm CO(17-16)}^{XDR}$) is the CO(17--16) flux predicted by PDR (XDR) models, and $\Omega_{OA}=4\pi(1-\cos\theta_T)$ is the solid angle subtended by the torus opening angle $\theta_T=35^{\rm o}$ (Lawrence et al. 1991).\\
Currently, the deepest X-ray survey reaches a flux threshold $F_{thr}= 9.1\times 10^{\rm-18}~\rm erg~s^{-1}cm^{-2}$ in 4 Ms of Chandra observing time (Xue et al. 2011), while the flux detection limit of $z\sim 6$ quasar observations typically is $\geq 3\times 10^{-15}\rm erg~s^{-1}~cm^{-2}$ (Shemmer et al. 2006; Page et al. 2013). In Fig. 3, we compare Chandra versus ALMA detectability of faint or obscured $(F_{\rm X}^{\rm obs} < 10^{-17}~\rm erg~s^{-1}cm^{-2}$) SMBH progenitors for different properties of their host galaxy MCs. 
If such sources do exist, their CO(17--16) line emission will be easily detectable by ALMA.
For example, for a typical case in which $n_{\rm XDR}\sim 10^6$ and $F_X\sim 28$ (magenta diamond in Fig. 3), our model predicts $L_{\rm CO(17-16)}\sim 10^8~ L_{\odot}$ and $F_{\rm X}^{\rm obs}\sim 4\times 10^{-18}~\rm erg~s^{-1}cm^{-2}$. Hence, while this source is detectable with the ALMA full array in BAND 6 (211$<\nu_{\rm obs}/{\rm GHz} < 275$) in 2.6 hr of observing time, its X-ray flux would remain unseen even in the deepest Chandra observations obtained so far. This is because X-rays photons remain trapped in dense molecular gas, where they are efficiently reprocessed into lower energy photons and particles heating the gas and efficiently boosting the CO(17--16) line to detectable luminosities. 
\begin{figure}
   \begin{center} 
   \psfig{figure=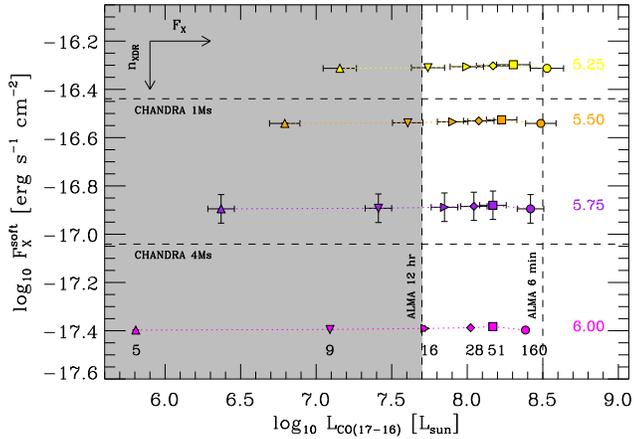,width=85mm,clip=t,angle=0}
   \end{center}
\caption{Comparison between the observed soft X-ray flux ($F_{\rm X}^{\rm soft}$) and CO(17--16) luminosity ($L_{\rm CO(17-16)}$) for a $z=7$ quasar, powered by a $3\times 10^{6}~\rm M_{\odot}$ black hole, radiating at Eddington luminosity, and with an obscuring gas column density $N_{\rm H}>N_{\rm H}^*=10^{24}$ cm$^{-2}$. Symbols and colors corresponds to different properties of the host galaxy ISM: yellow, orange, violet, magenta and red symbols refer to increasing XDR densities in the range $10^{5.25}<n_{\rm XDR}/\rm cm^{-3}<10^{6}$; triangles, downwards triangles, rightwards triangles, diamonds, squares, circles denote increasing X-ray fluxes at the illuminated surface of the XDR in the range $5<F_X<160 ~\rm erg~s^{-1}cm^{-2}$. Errorbars take into account the contribution of different PDR densities ($10^3<n_{\rm PDR}/\rm cm^{-3}<10^{3.5}$) to the $L_{\rm CO(17-16)}$ luminosity. We emphasize that the intrinsic X-ray flux of our sample source is fixed, and related to its bolometric luminosity through bolometric corrections. Hence, different $F_X$ values correspond to different distances between the central engine and the molecular clouds. 
}
\label{fig3}
\end{figure} 
\section{Discussion}
Our detection and its interpretation outline exciting possibilities for future discoveries of early SMBH progenitors. Such searches will be enabled by ALMA and future facilities as NOEMA (NOrthern Extended Millimeter Array\footnote{http://iram-institute.org/medias/uploads/\\NoemaBrochureEngFinal.pdf}) and CCAT (Cerro Chajnantor Atacama Telescope, Woody et al. 2012). We defer to a future work a detailed assessment of the optimal observational strategy and/or instrumental design to detect SMBH ancestors through high-J CO lines. Here, we critically analyze the limits and uncertainties of our detection of the CO(17--16) line.\\
In Sec. 2, we have largely discussed the possible OH$^+$ contamination of the CO(17--16) line. First of all, we underline that the presence of $\rm OH^+$ in J1148 lends further support to the proposed interpretation, namely that a strong contribution from XDRs is required to explain our observations, since $\rm OH^+$ emission/absorption is often ascribed to XDRs (van der Werk et al. 2010; Gonzalez-Alfonso et al. 2013).\\ 
We further note that our data require the XDR contribution to the observed COSLED as far as the OH$^+$ contribution to the serendipitous emission we detected is $<$30\%. An higher contribution is disfavored by the double fit analysis, since the OH$^+$ transition dominates the emission only if FWHM$_{\rm OH^+}\geq 500$\,km\,s$^{-1}$, i.e. for FWHM values much larger than all the other FIR lines detected so far in J1148. Clearly, to definitively establish the OH$^+$ and CO(17--16) relative contribution deeper observations are required.\\
We have also examined the possibility that the observed COSLED can instead result from shock excitation of the gas (Meijerink et al. 2013; Hailey-Dunsheath et al. 2012; Panuzzo et al. 2010). We can safely exclude this hypothesis based on the following arguments. In Fig. 2, we compare the COSLED of J1148 with those of three well known shock-dominated sources, typically starburst and Seyfert galaxies (as M82 or NGC1068) or merger systems (NGC6240). None of the sources shows the large high-$J$ line fluxes observed in J1148 and their COSLED appears exceedingly flat. In addition, a key difference between shocks and radiative excitation by UV and/or X-rays is that shocks only heat the gas without affecting the temperature of dust grains (Meijerink et al. 2013). As dust is the primary source of IR emission, if excitation were due to shocks one would expect a large ratio of the CO to IR luminosity. This is indeed measured in NGC6240, which has an unusually large $L_{CO}/L_{IR}=7\times 10^{-4}$ ratio, obtained by considering all CO transitions up to the $J=13-12$ one. For J1148, the rest-frame far infrared spectral energy distribution due to dust emission is extremely well sampled (Bertoldi et al. 2003b, Robson et al. 2004, Beelen et al. 2006), allowing for a precise measurement of the high ($L_{FIR}=2.0\pm 0.3\times 10^{13}~L_{\odot}$) FIR luminosity in this source (Valiante et al. 2011). 
We integrate the best-fit COSLED up to the $J=13-12$ transition for a fair comparison with NGC6240. We obtain $L_{CO}/L_{FIR}\approx 10^{-4}$, which is $\approx$7 times lower than the one found in NGC6240. Thus, we conclude that it is very unlikely that the CO(17-16) line of J1148 is powered by shocks. As a corollary, given the characteristic shape of the XDR COSLED, observing a prominent $J\sim 15-20$ line in high redshift objects would lend strong support to the presence of an X-ray source. In fact, UV radiation is unable to substantially power lines with $J \gsim 10$, and powerful, pervasive shocks affecting the large molecular masses in place would be required.\\
We conclude that, in spite of some uncertainties, our detection of the CO(17--16) line is robust and possibly opens a new pathway to search for the elusive SMBH progenitors. Further support to our interpretation should come from the detection of other high-J CO lines in J1148 and in other $z>6$ quasars. 
\section*{Acknowledgments}
SG thanks INAF for support through an International Post-Doctoral Fellowship. The authors are grateful to the anonymous referees for their comments that have strengthened the results presented. We thank C. Puzzarini, V. Lattanzi, E. Piconcelli and L. Zappacosta for useful comments. We acknowledge discussions with L. Vallini, L. Hunt, D. Lutz, R. Giovannelli, G. Stacey, X. Fan, E. Komatsu, A. Maselli. This work is based on observations carried out with the IRAM Plateau de Bure Interferometer. IRAM is supported by INSU/CNRS (France), MPG (Germany) and IGN (Spain).\\

\noindent
{\bf REFERENCES}\\

\noindent
Ao, Y., Weiss, A., Downes, D., Walter, F., Henkel, C., Menten, K.~M., 2008, A\&A, 491, 747--754\\
Barvainis, R., Maloney, P., Antonucci, R., \& Alloin, D. 1997, ApJ, 484, 695\\
Beelen, A., et al., 2006, ApJ, 642, 694--701\\
Bertoldi, F., et al., 2003a, A\&A, 409, L47--L50\\
Bertoldi, F., Carilli, C.L., Cox, P., Fan, X., Strauss, M.A., Beelen, A., Omont, A. \& Zylka, R., 2003b, A\&A, 406, L55--L58\\
Casey, C.~M., Narayanan, D., Cooray, A., 2014, Physics Reports, 514, 45--161\\
Comastri, A., 2004, ASSL, 308, 245--273\\
Cowie, L., et al., 2013, ApJ, 748, id.50\\
Fan, X. et al., 2003, AJ, 125, 1649--1659\\
Fan, X., 2006, NewAR, 50, 665\\
Ferrarese, L. \& Merritt, D., 2000, ApJ, 539, L9--L12\\
Gallerani, S. et al., 2012, A\&A, 543, A114\\
Gebhardt, K. et al., 2000, ApJ, 539, L13--16\\
Gonzalez-Alfonso, E. et al., 2013, A\&A, 550, 25--48, 2013\\
Greve, T.~R. et al., 2014, arXiv:1407.4400, accepted for publication in ApJ\\
Hailey-Dunsheath et al., 2012, ApJ, 755, 55--74\\
Hopkins P.~F., et al., 2007, ApJ, 654, 731\\ 
Jiang, L., 2006, AJ, 132, 2127\\
Juarez, Y.,  et al., 2009, A\&A, 494, L25--L28\\
Kurk, J.~D., Walter, F., Fan, X., Jiang, L., Jester, S., Rix, H.-W., \& Riechers, D. A. 2009, ApJ, 702, 833\\
Kurk, J. D., et al. 2007, ApJ, 669, 32\\
Lawrence, A., et al., 1991, MNRAS, 252, 586--592\\
Maiolino, R., et al., 2005, A\&A, 440, L51--L54\\
Maiolino, R., et al., 2012, MNRAS, 425, L66--L70\\
Meijerink, R. \& Spaans, M., 2005, A\&A, 436, 397--409\\
Meijerink, R., Spaans, M., Israel, F.~P., 2007, A\&A, 461, 793--811\\
Meijerink, R., et al., 2013, ApJL, 762, L16--L20\\
Morrison, R. \& McCammon, D., 1983, ApJ, 270, 119--122\\
Mortlock, D.~J., et al., 2011, Nature, 474, 616--619\\
Neri, R., et al., 2014, arXiv:1401.2396\\  
Nagao, T., et al., 2012, A\& A, 542, L34\\
Obreschkow, D., et al., 2009, ApJ, 702, 1321--1335\\
Omont, A., Petitjean, P., Guilloteau, S., McMahon, R. G., Solomon, P. M., Pecontal, E., 1996, Nature, 382, 428\\
Page, M.~J., et al., 2013, arXiv:1311.1686\\
Panuzzo, P., et al., 2010, A\&A, 518, L37--L41\\
Petri, A., Ferrara, A. \& Salvaterra, R. 2012, MNRAS, 422, 1690\\
Planck Collaboration, et al., 2013, arXiv:1303.5076v1\\
Riechers, D., et al., 2009, ApJ, 703, 1338--1345\\
Riechers, D., et al., 2013, Nature, 496, 329\\
Risaliti, G. \& Elvis, M., 2004, ASSL, 308, 187\\
Robson, I., Priddey, R.S., Isaak, K.G., McMahon, R.G., 2004, MNRAS, 351, L29--L33\\
Schleicher D., et al., 2010, A\&A, 513, 7--21\\
Shemmer, O., et al., 2006, ApJ, 644, 86--99\\
Solomon, P.M., et al., 1992, Nature, 356, 318--321\\
Telfer, R.C., et al., 2002, ApJ, 565, 773--785\\
Tanaka, T. \& Haiman, Z., 2009, ApJ, 696, 1798--1822\\
Valiante, R. et al., 2011, MNRAS, 416, 1916--1935\\ 
van der Werk, P.~P. et al., 2010, A\&A, 518, L42--L46\\
Volonteri, M., 2010 A\&AR, 18, 279--315\\
Xue, Y.~Q., et al., 2011, ApJS, 195, 10--41\\
Walter, F., et al., 2003, Nature, 424, 406--408\\
Walter, F., et al., 2009, Nature, 457, 699--701\\
Wang, R., et al. 2010, ApJ, 714, 699\\
Weiss A., et al., 2007, A\&A, 467, 955\\
White, R.L., et al., 2003, AJ, 126, 1--14\\
Willott, C.~J., et al., 2003, ApJ, 587, L15--L18\\
Woody, D., et al., 2012, Proceedings of the SPIE, 8444, 14
\end{document}